\documentclass{article}
\usepackage[T1]{fontenc}

\makeatletter

\providecommand{\LyX}{L\kern-.1667em\lower.25em\hbox{Y}\kern-.125emX\@}

\newcommand{\lyxaddress}[1]{
  \par {\raggedright #1 
  \vspace{1.4em}
  \noindent\par}
}

\usepackage[T1]{fontenc}
\usepackage{geometry}
\usepackage{color}
\usepackage{setspace}

\begin{document}

\title{GREEN'S FUNCTIONS IN AXIAL AND LORENTZ-TYPE GAUGES AND APPLICATION TO THE AXIAL
POLE PRESCRIPTION AND THE WILSON LOOP\thanks{
{*}talk presented at the workshop QFT2000,held at Saha Institute of Nuclear
Physics , Calcutta,INDIA from 18 Jan 2000 till 22 Jan 2000. 
},\thanks{
email address:sdj@iitk.ac.in 
}}

\author{Satish D. Joglekar}

\maketitle

\lyxaddress{Department of Physics, I.I.T.Kanpur,Kanpur 208016{[}INDIA{]}}

We summarize the work done in connecting Green's functions in a different classes
of gauges and its applications to the problems in the axial gauges.The procedure
adopted uses finite field-dependent BRS {[}FFBRS{]} transformations to connect
axial and the Lorentz type gauges.These transformations preserve the vacuum
expectation of gauge-invariant observables explicitly. We discuss the applications
of these ideas to the axial gauge pole problem and to the preservation of the
Wilson loop and the thermal Wilson loop.

\section{MOTIVATIONS FOR THE WORK}

This work summarizes the progress made recently in applying the finite field-dependent
BRS {[}FFBRS{]} transformations to connect the axial and the Lorentz type gauges
and the resultant applications to the problems associated with the axial gauges.These
transformations preserve the vacuum expectation of gauge-invariant observables
by their explicit construction.The summary is based on the works {[}1-4{]} with
Aalok Misra and on {[}5{]} and is based on the earlier works {[}6,7{]}.

~

The known high energy physics is well represented by the standard model ;which
is a non-abelian gauge theory. Hence~ practical~ calculations~ (and~ their~
methodology)~ in~ nonabelian gauge theories assume importance in Particle~ Physics.
Practical calculations in a non-abelian gauge~ theory~ require~ a~ choice of
a gauge and there are many choices available. {[}Theories involving gravity
also has a gauge-invariance and require a gauge choice.{]} There are many distinct
{[}families of{]} gauges, variously useful~ in different situations. For example,
we have:

{[}1{]} Lorentz Gauges:~ 

{[}2{]}Coulomb Gauge: 

{[}3{]}Axial Gauges:~~~~ {[}Includes Light-cone,Temporal etc.{]}

{[}4{]}Planar Gauges:

{[}5{]}radial Gauges:

{[}6{]}Quadratic Gauges:

{[}7{]}in SBGT R-xi Gauges;etc.

The uses of these gauges in a variety of different contexts have been elaborated
in reference 8. A priori, we expect from gauge invariance,that the values for
physical observables calculated in different gauges are identical. Formal proofs
of such equivalence for the S-matrix elements has been given in a given class
of gauges, say the Lorentz-type gauges with a variable gauge parameter \( \lambda  \)
{[}9,10{]}.Some isolated attempts to connect S-matrix elements in singular {[}rather
than a class of them{]} gauges also have been done. For example, formal equivalence
of S-matrix elements in the Coulomb and the Landau gauges {[}both singular gauges{]}
has also been established {[}11{]}.Similar formal attempts to connect the {[}singular{]}
temporal gauge with Feynman gauge in the canonical formalism has also been done
{[}12,13{]}. It is important to note that, however, the Green's functions in
the gauges such as the Coulomb {[}14{]},the axial,the planar and the light-cone
{[}15,16{]} in the path integral formulation are ambiguous on account of the
unphysical singularities in their propagators. Hence, it becomes important to
know how to define the Green's functions in such gauges in such a manner that
they are compatible to those in a well-defined covariant gauge such as the Lorentz
gauge. A general procedure that connects Green's functions in the path integral
formulation in two classes of gauges, say the Lorentz to the axial,has been
lacking until recently. Such comparisons are important not just in a formal
sense but also in practice.Precisely because of this, the proper treatment of
the 1/\( (\eta .q)^{p} \) type poles in axial and light-cone gauges (and also
similar questions in the Coulomb gauge {[}14{]}) has occupied a lot of attention
{[}15,16{]} and the criterion used for their validation has, in fact, been the
comparison with the calculational results in the Lorentz gauges.Such comparative
calculations, where possible {[}15,16{]} have to be done by brute force and
have been done to O{[}g\( ^{4} \){]} generally; thus limiting the scope of
their confirmation. At a time, a physically observable anomalous dimension was
reported to differ in Lorentz and axial gauges {[}8{]}. Such questions motivate
us to develop a general path integral formalism that can address all these questions
in a wide class of gauges in a single framework. In a purely Feynman diagrammatic
approach,we ,of course, have the attempt of Cheng and Tsai{[}14 {]}.

Further, gauges such as axial gauges suffer from the prescription ambiguity
for the 1/(n.q)\( ^{p} \) type singularities. Several ad hoc prescriptions
have been given; of these the Principal value prescription{[}PVP{]} and the
Leibbrandt-Mandelstam{[}M-L{]} are the ones used extensively in literature.While
much work has been done on these,they however have run inevitably into difficulties
of various kinds {[}19,15{]}.Such ambiguities do not exist in Lorentz type gauges.
We thus expect that a field transformation from the Lorentz gauges to axial
gauges will enable us to derive the correct prescription for such singularities.

\section{THE PROCEDURE ADOPTED}

Such considerations motivate us to seek a field transformation that relates
different family of gauges. In this section,we shall outline in brief the procedure
we adopt in these works.The details of the procedure will be left to the next
section.The procedure consists of :

\begin{itemize}
\item {[}1{]}Construct a field transformation that~~ transforms the path integral
from one gauge choice to~ another.
\item {[}2{]} Develop method to correlate the Green's functions in two sets of gauges~
formally.
\item {[}3{]} Establish a relation between propagators. This is expected to give an
understanding of how to deal with~ 1/n.q~ -type singularities in the axial gauge
propagator.{[}The prescription problem.{]}
\end{itemize}
We know explicitly the infinitesimal gauge transformation {[}9{]} that relates
two gauge functions differing infinitesimally; viz.

~~~~~~~~~~~~~~ F\( ^{\alpha } \) -->~ F\( ^{\alpha } \) + \( \delta  \)F\( ^{\alpha } \) 

It is given by

~~~~~~~~~~~~~~~~~~~ \( \delta  \)A\( ^{\alpha } \)\( _{\mu } \) = D\( ^{\alpha \beta }_{\mu } \)
\( M^{-1}_{\beta \gamma }\delta  \)F\( ^{\gamma } \)

This is an example of a field-dependent infinitesimal gauge transformation.It
is difficult to integrate and explicitly evaluate the finite version of this.For
example,even in a simple case such as seeking the transformation from~ \( \partial  \).A
= 0 {[}Landau{]} to A\( _{0} \)'= 0{[}temporal{]} gauge,the transformation
can be formally solved for ,but is difficult to evaluate explicitly. Here,A'
is given by

A' = \( \frac{1}{g} \)U{[}A{]} {[} \( \partial  \)+gA{]}U{[}A{]}\( ^{-1} \)

With~

U{[}A{]} = T \{exp \( \int  \)\( _{_{0}}^{t} \) A\( _{0} \)(x,t') dt'\}

For such reasons, we instead seek an alternate approach in which we try to integrate
an analogue of the BRS transformations.This seems to be more easily manageable.
This property arises mainly from the facts that{[}1{]} BRS transformations are
nilpotent, {[}2{]} The ``finite'' BRS and ``infinitesimal'' BRS have the
same form unlike the gauge transformations\footnote{
In the context of the usual BRS transformations,where ``\( \delta \Lambda  \)''
and ``\( \Lambda  \)'' are merely \emph{constants,}this remark may look trivial
because \( \Lambda ^{2} \)=0;however in connection with the FFBRS transformations,this
distinction is not trivial.See the remark below (3.3). 
} .

\section{FFBRS TRANSFORMATIONS:A GENERALIZATION OF BRS}

~We note the invariance of the Faddeev-Popov Effective Action {[}FPEA{]} in
a gauge 'F' is not dependent on the anticommuting parameter ``\( \delta \Lambda  \)''
being infinitesimal nor on whether it is field-independent or not {[}6{]}.In
fact transformations such as these:

A'(x) = A(x) + D\( _{\mu } \)c(x) \( \Theta  \){[}A(y),c(y),\( \overline{c} \)
(y){]}~~~~~~~~~~~~~~~~~~~~~~~~~~~~~~~~~~~~~~~~~~~~~~~~~~~~~~~~~~~~~~~(3.1a)

c\( ^{\alpha } \)'(x) = c\( ^{\alpha } \)(x)~ -- 1/2 g f\( ^{\alpha \beta \gamma }c^{\beta }c^{\gamma } \)
\( \Theta  \){[}A(y),c(y),\( \overline{c} \) (y){]}~~~~~~~~~~~~~~~~~~~~~~~~~~~~~~~~~~~~~~~~~~~~~~(3.1b)

\( \overline{c} \)'\( ^{\alpha } \)(x) = \( \overline{c} \)\( ^{\alpha } \)(x)
+ \( \frac{\partial \cdot A^{\alpha }}{\lambda } \) \( \Theta  \){[}A(y),c(y),\( \overline{c} \)
(y){]}~~~~~~~~~~~~~~~~~~~~~~~~~~~~~~~~~~~~~ ~~~~~~~~~~~~~~~~~~~(3.1c)

or in brief,

\( \phi _{i}' \)=\( \phi _{i} \) +\( \delta _{i,BRS}[\phi  \){]}\( \Theta  \){[}A(y),c(y),\( \overline{c} \)
(y){]}~~~~~~~~~~~~~~~~~~~~~~~~~~~~~~~~~~~~~~~~~~~~~~~~~(3.2)

where \( \Theta  \){[}A(y),c(y),\( \overline{c} \) (y){]} is a field-dependent
functional but does not depend on 'x' are also symmetries of the FPEA.\( \Theta  \)
need not be ``infinitesimal'',in that the Green's functions of \( \Theta  \)
with appropriate external lines can be a finite number even when \( \Theta  \)\( ^{2} \)=0.For
examples of these, see e.g. Ref.8 and some of these will also appear in applications
below.

It has been shown{[}6{]} that in some special cases of these,the corresponding
FFBRS can be obtained by integration of~ an infinitesimal field-dependent BRS
{[}IFBRS{]} transformation: \(  \)

~ d\( \phi  \)(x,\( \kappa  \))/d\( \kappa  \) =\( \delta _{BRS} \) {[}\( \phi  \)(x,\( \kappa  \))){]}\( \Theta ' \)
{[}\( \phi  \)(y,\( \kappa  \)){]}~~~~~~~~~~~~~~~~~~~~~~~~~~~~~~~~~~~~~~~~~~~~~(3.3)

which stand for the three equations:

dA(x,\( \kappa  \))/d\( \kappa  \)=D\( _{\mu } \)c(x,\( \kappa  \)) \( \Theta  \)'{[}A(y,\( \kappa  \)),c(y,\( \kappa  \)),\( \overline{c} \)
(y,\( \kappa  \)){]}~~~~~~~~~~~~~~~~~~~~~~~~~~~~~~~~~~~~~~~~~~(3.4a)

dc(x,\( \kappa  \))/d\( \kappa  \)=- 1/2 g f\( ^{\alpha \beta \gamma }c^{\beta }(x,\kappa )c^{\gamma } \)
(x,\( \kappa ) \)\( \Theta  \)'{[}A(y,\( \kappa  \)),c(y,\( \kappa  \)),\( \overline{c} \)
(y,\( \kappa  \)){]}~~~~~~~~~~~~~~~~~~~~~~~~~(3.4b) 

d\( \overline{c} \)(x,\( \kappa  \))/d\( \kappa  \)= \( \partial  \)\( \cdot  \)A\( ^{\alpha } \)(x,\( \kappa ) \)/\( \lambda  \)
\( \Theta  \)'{[}A(y,\( \kappa  \)),c(y,\( \kappa  \)),\( \overline{c} \)
(y,\( \kappa  \)){]} ~~~~~~~~~~~~~~~~~~~~~~~~~~~~~~~~~~~~(3.4c)

In this case, the integral can be evaluated in a closed form.When integrated
it, in fact, maintains its BRS form{[}6{]} on account of the two properties
of BRS mentioned earlier.\{For a more transparent derivation,you may also refer
to {[}8{]}\}.This is much simpler than and unlike a gauge transformation where
a finite gauge transformation does not retain the simple form of an infinitesimal
gauge transformation.The result,thus,exhibits the FFBRS structure of (3.1),
with \( \Theta  \) given by:

~~~~~~~~~~~~ \( \Theta  \){[}\( \phi  \){]} = \( \Theta  \)'{[}\( \phi  \){]}~
\{exp(f{[}\( \phi  \){]}) - 1\}/ f{[}\( \phi  \){]}~~~~~~~~~~~~~~~~~~~~~~~~~~~~~~~(3.5) 

with~ 

f =\( \sum  \)\( _{_{i}} \) {[}\( \delta  \)\( \Theta  \)'/d\( \phi _{i} \){]}
\( \delta _{i,BRS} \) {[}\( \phi  \)(x,\( \kappa  \))){]}~~~~~~~~~~~~~~~~~~~~~~~~~~~~~~~~~~~~~~~~~~~~~~~~~(3.6) 

The Jacobian for the above non-local transformation is difficult to evaluate
directly.However,it turns out that in many cases of interest, the Jacobian for
such FFBRS can effectively be cast in the form exp \{iS\( _{1} \)\}with a local
S\( _{1} \);which in fact leads to a total effective action \{S\( _{eff} \)+S\( _{1} \)\}
that is an FPEA for another gauge theory {[}6{]}.Thus {[}in such cases{]}, the
FFBRS converts one path-integral\footnote{
for a more accurate statement, please see reference 6. 
} into another rather than one action to another;the action being in fact invariant
under under the FFBRS.To be explicit,if we express D\( \phi  \)=D\( \phi  \)'J{[}\( \phi  \)'{]}.Then
,in such cases, for any gauge-invariant observable G{[}\( \phi ] \),we have

<\textcompwordmark{}<G{[}\( \phi  \){]}>\textcompwordmark{}>\( _{L} \)= \( \int  \)D\( \phi  \)
G{[}\( \phi  \){]}exp\{ i S\( _{eff} \)\( ^{L}[ \)\( \phi  \)\( ] \)\}=\( \int  \)D\( \phi  \)'J{[}
\( \phi  \)'{]}G{[}\( \phi ' \){]}exp\{ i S\( _{eff} \)\( ^{L}[ \)\( \phi ' \)\( ] \)\}

\( \equiv  \)\( \int  \)D\( \phi  \)'G{[}\( \phi ' \){]}exp\{ i S\( _{eff} \)\( ^{L}[ \)\( \phi ' \)\( ] \)+iS\( _{1}[ \)\( \phi ' \){]}\}=\( \int  \)D\( \phi  \)'G{[}\( \phi ' \){]}exp\{
i S\( _{eff} \)\( ^{A}[ \)\( \phi ' \)\( ] \)\}~~~~~~~~~~~~~~~~~~~~~(3.7) 

Here, S\( _{eff} \)\( ^{A} \) is the effective action of another gauge such
as the axial;

In particular,for G{[}\( \phi ] \)= I,we have that \( \phi  \)\( \rightarrow  \)\( \phi  \)'
transforms vacuum to vacuum amplitude in the two gauges.

We thus outline a general prescription for constructing an FFBRS connecting
any two families of Yang-Mills effective actions: 

\begin{itemize}
\item {[}a{]} Establish a continuous route of interpolating gauges {[}if necessary{]}
from one family to another;
\item {[}b{]}Postulate an infinitesimal field-dependent BRS transformation.The form
of the infinitesimal gauge transformation (where available) serves as a preliminary
hint.
\item {[}c{]} Using the form for the interpolating S, if necessary, guess a form for
S\( _{1} \){[}\( \phi  \),\( \kappa  \){]}.
\item {[}d{]} Evaluate the Jacobian for an infinitesimal BRS in step {[}b{]}.This
is easily evaluated compared to that for an FFBRS.
\item {[}e{]}Impose the condition meant for J\( \rightarrow  \)exp\{ iS\( _{1} \)
\}. This condition leads to constraints on \( \Theta  \)' and the coefficients
and the form for S\( _{1} \) .This condition reads {[}6{]}:
\end{itemize}

\subparagraph*{<\textcompwordmark{}<\protect\( \frac{i}{J}\protect \)\protect\( \frac{dJ}{d\kappa }\protect \)+\protect\( \frac{dS_{1}[\phi (\kappa ),\kappa ]}{d\kappa }\protect \)>\protect\( _{\kappa }\protect \)\protect\( \equiv \protect \)0
~ ~~~~~~~~~~~~~~~~~~~~~~~~~~~~~~~~~~~~~~~~~~~~~~~~~~~~~~~(3.8)}

\subparagraph*{and involves the Jacobian for the infinitesimal ~transformation of (3.4)which
is easy to evaluate.~ }

\paragraph*{We note that while the above procedure is required while dealing with arbitrary
effective actions for Yang-Mills theories, in the special case of arbitrary
two Faddeev-Popov Effective Actions {[}FPEA{]}, a simpler proof has also recently
been given{[}17{]}. As examples of the former class, not included in the latter,
we note, for clarity, that the latter class does not include BRS-antiBRS effective
action of Baulieu and Thierry-Mieg as well as say the BRS invariant action in
planar gauges. For such cases we need to follow the entire process as presented
{[}6,17{]}.}

\section{THE RESULTS FOR THE FFBRS FOR LORENTZ \protect\( \rightarrow \protect \)AXIAL}

We follow the procedure as outlined in section 3.{[}Alternately we can also
use the result of ref.17 ,now available.{]}

First we understand the n.A=0 gauge as the \( \lambda \rightarrow 0 \) limit
of the gauge with

~ S\( _{gf} \) = -\( \frac{1}{2\lambda }\int  \) d\( ^{4}x \) {[}n.A{]}\( ^{2} \)~~~~~~~~~~~~~~~~~~~~~~~~~~~~~~~~~~~~~~~~~~~~~~~~~~~~~~~~~(4.1)

(~{[}together with the corresponding ghost term{]}.

Then we construct an intermediate gauge-fixing term

S\( _{gf} \) = -\( \frac{1}{2\lambda }\int  \) d\( ^{4}x \) {[}(1- \( \kappa  \))\( \partial \cdot  \)A
+\( \kappa  \) n.A{]}\( ^{2} \) ~~~~~~~~~~~~~~~~~~~~~~~~~~~~~~~~~~~~~~~~~~~~~~~~~~~(4.2) 

together with the corresponding ghost term. From these,we make an ansatz for
\( \Theta  \)' and S\( _{1} \); and impose the Jacobian condition.We then
obtain as one possible solution the following:

~~~ \( \Theta  \)'{[}\( \phi  \)(\( \kappa  \)){]} =\( i\int d^{4}y \) \( \overline{c} \)\( ^{\gamma } \)(y,\( \kappa  \))
{[}\( \partial . \)A\( ^{\gamma } \)(y,\( \kappa ) \) -\( \eta  \).A\( ^{\gamma } \)(y,\( \kappa )] \)~~~~~~~~~~~~~~~~~~~~~~~~~~~~~~~~~~~~~~~~~~~~~~~~~~~(4.3)

Then,the FFBRS transformation that takes one from the Lotentz-type gauges to
the axial-type gauges is given by (3.1),with \( \Theta  \) given by

~~~~~~~~~~ \( \Theta  \){[}\( \phi  \){]} = \( \Theta  \)'{[}\( \phi  \){]}~
\{exp(f{[}\( \phi  \){]}) - 1\}/ f~~~~~~~~~~~~~~~~~~~~~~~~~~~~~~~~~~~~~~~~~~~~~~~~~~~(4.4)

and 

f{[}\( \phi  \){]}=\( \sum  \)\( _{_{i}} \) {[}\( \delta  \)\( \Theta  \)'{[}\( \phi  \){]}
/d\( \phi _{i} \){]} \( \delta _{i,BRS} \) {[}\( \phi  \)(x){]}~~~~~~~~~~~~~~~~~~~~~~~~~~~~~~~~~~~~~~~~~~~~~~~~~~~(4.5a) 

~~~~~=\( i\int d^{4}y \) \{\( \overline{c} \)\( ^{} \)(y){[}\( \partial  \)-\( \eta  \){]}
D\( _{\mu } \)c(y) + \( \partial . \)A\( ^{\gamma } \)(y\( ) \){[}\( \partial . \)A\( ^{\gamma } \)(y)
-\( \eta  \).A\( ^{\gamma } \)(y){]}\( \frac{1}{\lambda } \)\}~~~~~~~~~~~~~(4.5b)

We mention some of the properties of the results:

{[}A{]} While the results look complicated, a simple formula that enables one
to evaluate the Green's functions in one set of gauges in terms of the Feynman
rules in another set of gauges is possible~ as seen in the next section.

{[}B{]}Similar procedure can be followed for other pairs of gauges not otherwise
connected.{[}See references 6 and 17{]}.

{[}C{]}The transformations preserve the vacuum expectation values of gauge invariant
operators automatically.e.g.The Wilson Loop {[}See section 8{]}.

\section{THE RESULTS FOR GREEN'S FUNCTIONS FOR THE TWO GAUGES.}

~

Unlike the path integral and the vacuum expectation value of gauge invariant
observables, the relation between Green's functions in the two sets of gauges
requires another, but related transformation {[}1{]}. This transformation turns
out to be an integral of the IFBRS {[}1{]}:

~dA(x,\( \kappa  \))/d\( \kappa  \)=D\( _{\mu } \)c(x,\( \kappa  \)) \( \Theta  \)'{[}A(y,\( \kappa  \)),c(y,\( \kappa  \)),\( \overline{c} \)
(y,\( \kappa  \)){]}~~~~~~~~~~~~~~~~~~~~~~~~~~~~~~~~~~~~~~~~~~~~~~~~~~~(5.1a)

dc(x,\( \kappa  \))/d\( \kappa  \)=- 1/2 g f\( ^{\alpha \beta \gamma }c^{\beta }(x,\kappa )c^{\gamma }(x,\kappa  \))
\( \Theta  \)'{[}A(y,\( \kappa  \)),c(y,\( \kappa  \)),\( \overline{c} \)
(y,\( \kappa  \)){]}~~~~~~~~~~~~~~~~~~~~~~~~~(5.1b) 

d\( \overline{c} \)(x,\( \kappa  \))/d\( \kappa  \)= \( [\partial  \)\( \bullet  \)A\( ^{\alpha }(1-\kappa )+\kappa \eta .A \)\( ^{\alpha }] \)/\( \lambda  \)
\( \Theta ' \){[}A(y,\( \kappa  \)),c(y,\( \kappa  \)),\( \overline{c} \)
(y,\( \kappa  \)){]}~~~~~~~~~~~~~~~~~~~~~(5.1c)

We can re-express these in a single equation:

d\( \phi  \)\( _{i} \)(x,\( \kappa  \))/d\( \kappa  \) \( \equiv  \)\( (\widetilde{\delta _{1i}[\phi } \){]}+\( \kappa  \)\( (\widetilde{\delta _{2i}[\phi } \){]})\( \Theta ' \){[}A(y,\( \kappa  \)),c(y,\( \kappa  \)),\( \overline{c} \)
(y,\( \kappa  \)){]}~~~~~~~~~~~~~~~~~~(5.2)

Let us say that the integral of the above from \( \kappa  \)=0 to ~\( \kappa  \)=1
reads

\( \phi ' \)\( \equiv  \)\( \phi (x,\kappa =1)\equiv  \) \( \phi (x,\kappa =0) \)\( +\Delta \phi [ \)\( \phi ]= \)\( \phi +\Delta \phi  \)~~~~~~~~~~~~~~~~~~~~~~~~~~~~~~~~~~~(5.3)

It is the above solution (5.3) that appears in the relation between the Green's
functions between two gauges.In fact, defining for a gauge 'F'

<\textcompwordmark{}<O{[}\( \phi  \){]}>\textcompwordmark{}>\( _{F} \)= \( \int  \)D\( \phi  \)
O{[}\( \phi  \){]}exp\{ i S\( _{eff} \)\( ^{F}[ \)\( \phi  \)\( ] \)\}~~~~~~~~~~~~~~~~~~~~~~~~~~~~~~~~~~~~~~~~~~~(5.4)

The result of this is that 

<\textcompwordmark{}<O{[}\( \phi  \){]}>\textcompwordmark{}>\( _{A} \) = <\textcompwordmark{}<
O {[}\( \phi +\Delta \phi  \){]}>\textcompwordmark{}>\( _{L} \)~~~~~~~~~~~~~~~~~~~~~~~~~~~~~~~~~~~~~~~(5.5)

\( \Delta \phi  \) can in fact be given through the relations {[}1,5{]}:

\( \Delta \phi  \) = \{\( \widetilde{\delta _{1}} \)\( [\phi ]\Theta _{1}[\phi ] \)\( +\widetilde{\delta _{2}} \)\( [\phi ]\Theta _{2}[\phi ] \)\}\( \Theta '[\phi ] \)~~~~~~~~~~~~~~~~~~~~~~~~~~~~~~~~~~~~(5.6)

where as defined in {[}1{]},

\( \Theta _{1,2}[\phi ] \)=\( \int  \)\( _{_{0}} \)\( ^{1} \)d\( \kappa  \)(1,\( \kappa  \))
exp \{\( \kappa f_{1}[\phi ]+\frac{\kappa ^{2}}{2}f_{2}[\phi ]\} \)~~~~~~~~~~~~~~~~~~~~~~~~~~(5.7)

The result again seems complicated, in principle, to be of use in evaluating
Green's functions; it can, however, be put in tractable form as below {[}1,5{]}\footnote{
See Equation (5.4) for definitions of \textcolor{magenta}{{\large \( \widetilde{\delta _{1i}[\phi } \){]}
and\( \widetilde{\delta _{2i}[\phi } \){]}.}}{\large \par{}} 
}:

<\textcompwordmark{}<O{[}\( \phi  \){]}>\textcompwordmark{}>\( _{A} \)'= <\textcompwordmark{}<O{[}\( \phi  \){]}>\textcompwordmark{}>\( _{L} \)
+i\( \int  \)\( _{_{0}} \)\( ^{1} \)d\( \kappa  \) D\( \phi  \) exp\{ i
S\( _{eff} \)\( ^{M}[ \)\( \phi  \),\( \kappa ] \)\}

~ ~~ ~~ ~~ ~~~ ~~ ~ ~~ ~~ ~~ ~~~ ~~~~ ~~ ~\( \bullet  \)\( \sum _{i}(\widetilde{\delta _{1i}[\phi } \){]}+\( \kappa  \)\( (\widetilde{\delta _{2i}[\phi } \){]})(-i\( \Theta ') \)\( \frac{\delta ^{L}O}{\delta \phi _{i}} \)~~
~ ~~ ~~ ~~~ (5.8) 

The proper definition of the Green's functions in Lorentz gauges requires that
we include a term -i\( \epsilon  \)\( \int  \)d\( ^{4}x \)(\( \frac{1}{2} \)AA
-\( \overline{c} \) c) in the effective action. Similarly,proper definition
of the axial Green's functions in Axial-type gauges requires that we include
an appropriate \( \epsilon  \)-term.The correct \( \epsilon  \)-term in axial-type
gauges can be obtained {[}2{]}by the FFBRS transformation and the relation of
the form (5.5)applied appropriately . It can be shown {[}2{]} that the effect
of this term on the axial gauge Green's functions is expressed in the simplest
form when it is expressed in the relation (5.8) and happens to be simply to
add to S\( _{eff} \)\( ^{M}[ \)\( \phi  \),\( \kappa ] \) the same -i\( \epsilon  \)\( \int  \)d\( ^{4}x \)(\( \frac{1}{2} \)AA
-\( \overline{c} \) c) inside the \( \kappa  \)-integral.Thus, taking care
of the proper definition of the axial Green's functions ,these Green's functions
,compatible with those in Lorentz gauges ,are given by

<\textcompwordmark{}<O{[}\( \phi  \){]}>\textcompwordmark{}>\( _{A} \)'= <\textcompwordmark{}<O{[}\( \phi  \){]}>\textcompwordmark{}>\( _{L} \)
+i\( \int  \)\( _{_{0}} \)\( ^{1} \)d\( \kappa  \) D\( \phi  \) exp\{ i
S\( _{eff} \)\( ^{M}[ \)\( \phi  \),\( \kappa ] \) +\( \epsilon  \)\( \int  \)d\( ^{4}x \)(\( \frac{1}{2} \)AA
-\( \overline{c} \) c)\}

~~ ~ ~~~ ~~ ~ ~~ ~~~ ~ ~~~ ~~ ~ ~~ ~~ ~~ ~\( \bullet  \)\( \sum _{i}(\widetilde{\delta _{1i}[\phi } \){]}+\( \kappa  \)\( (\widetilde{\delta _{2i}[\phi } \){]})(-i\( \Theta ') \)\( \frac{\delta ^{L}O}{\delta \phi _{i}} \)~~
~ ~~~ ~~ ~ ~~ ~~ ~~ ~~~ ~~~~ (5.9)

As seen above, a given Green's function~ to a given finite order can be evaluated
by means of a finite set of diagrams with vertices from the Lorentz gauges and
BRS variations AND the propagators from the mixed gauges.A \( \kappa  \)-integral
is also required to be performed.

An example:

Consider O{[}\( \phi  \){]}= A\( ^{\alpha }_{\mu }(x) \)A\( ^{\beta } \)\( _{\nu }(y). \)Then,
<\textcompwordmark{}<A\( ^{\alpha }_{\mu }(x) \)A\( ^{\beta } \)\( _{\nu }(y) \)>\textcompwordmark{}>\( _{A} \)
= i G\( ^{A\alpha \beta } \)\( _{\mu \nu } \)(x-y) for the connected part
of the axial gauge propagator.Then, in obvious notations, (5.9) reads,

iG\( ^{A\alpha \beta } \)\( _{\mu \nu } \)(x-y)=iG\( ^{L\alpha \beta } \)\( _{\mu \nu } \)(x-y)+i\( \int  \)\( _{_{0}} \)\( ^{1} \)d\( \kappa  \)
D\( \phi  \) exp\{ i S\( _{eff} \)\( ^{M}[ \)\( \phi  \),\( \kappa ] \)
+\( \epsilon  \)\( \int  \)d\( ^{4}x \)(\( \frac{1}{2} \)AA -\( \overline{c} \)
c)\}

~~ ~ ~~~ ~~ ~ ~~ ~~ ~ ~~ ~~ ~~ ~\( \bullet [ \)D\( _{\mu } \)c\( ^{\alpha } \)(x)A\( ^{\beta } \)\( _{\nu }(y) \)+A\( ^{\alpha }_{\mu }(x) \)D\( _{\mu } \)c\( ^{\beta } \)(y){]}
\( \int d^{4}z \) \( \overline{c} \)\( ^{\gamma } \)(z) {[}\( \partial . \)A\( ^{\gamma } \)(\( z) \)
-\( \eta  \).A\( ^{\gamma } \)(z){]}~~ ~~~ ~~~~ (5.10)

The above relation gives the value of the exact axial propagator compatible
to the Green's functions in Lorentz gauges .The result is exact to all orders.As
mentioned earlier, to any finite order in g,the right hand side can be evaluated
by a finite sum of Feynman diagrams. 

We can consider the above relation to O{[}g\( ^{0}] \).Then it will give the
free axial propagator compatible with the Lorentz gauges.This is automatically
expected to give information as to how the 1/n.q~ -type singularities in the
axial gauge propagator should be interpreted.{[}The prescription problem.{]}.It
reads,

iG\( ^{0A\alpha \beta } \)\( _{\mu \nu } \)(x-y)=iG\( ^{0L\alpha \beta } \)\( _{\mu \nu } \)(x-y)+i\( \int  \)\( _{_{0}} \)\( ^{1} \)d\( \kappa  \)
D\( \phi  \) exp\{ i S\( _{eff} \)\( ^{M}[ \)\( \phi  \),\( \kappa ] \)
+\( \epsilon  \)\( \int  \)d\( ^{4}x \)(\( \frac{1}{2} \)AA -\( \overline{c} \)
c)\}

~~ ~ ~~~ ~~ ~ ~~ ~~ ~ ~~ ~~ ~~ ~\( \bullet [ \)\( \partial  \)\( _{\mu } \)c\( ^{\alpha } \)(x)A\( ^{\beta } \)\( _{\nu }(y) \)+A\( ^{\alpha }_{\mu }(x) \)\( \partial _{\nu } \)c\( ^{\beta } \)(y){]}
\( \int d^{4}z \) \( \overline{c} \)\( ^{\gamma } \)(z) {[}\( \partial . \)A\( ^{\gamma } \)(\( z) \)
-\( \eta  \).A\( ^{\gamma } \)(z){]}~~ ~~~~ (5.11)

We, of course, need to evaluate the last term to O{[}g\( ^{0}] \). G\( ^{0} \)
refers to free propagator.This is discussed further in Sec.7.

\section{AN ALTERNATE DERIVATION BASED ON BRS:}

The result (5.8 )can also be given a direct derivation based on BRS{[}5{]}.Since,
for many calculations in practice, the latter result is sufficient, we give~
a glance of the procedure.

We write expectation of O{[}\( \phi  \){]} for the mixed gauge:

<\textcompwordmark{}<O{[}\( \phi  \){]}>\textcompwordmark{}>\( _{\kappa } \)\( =\int  \)
D\( \phi  \)O{[}\( \phi  \){]}exp\{ i S\( _{eff} \)\( ^{M}[ \)\( \phi  \),\( \kappa ] \)\}~~
~~~~ ~~ ~~~~ ~~ ~~~~~~ ~~ ~~~~~ ~~~~(6.1) 

~And evaluate d/d\( \kappa  \) of the above.We simplify the result for \( \frac{d}{d\kappa } \)<\textcompwordmark{}<O{[}\( \phi  \){]}>\textcompwordmark{}>\( _{\kappa } \)
using ,in particular,the BRS WT-identity of the Mixed gauge.(For the complete
technical details,please see {[}5,17{]}). We then integrate the result over
\( \kappa  \) from 0 to 1 in the end.This leads us to the equation (5.8).

We note ,however,as remarked earlier,that a complete definition of the Lorentz/Axial
gauge Green's function <\textcompwordmark{}<O{[}\( \phi  \){]}>\textcompwordmark{}>
requires an appropriate \( \epsilon  \)-term in each case as in equation (5.9
).The derivation of (5.9 )in reference 1, however,did require the use of FFBRS/IFBRS
construction in sections 4 and 5 explicitly.

\section{APPLICATION TO THE 1/n.q -type SINGULATITIES}

~

The equation (5.11 ), in the momentum space, reads:

G\( ^{0A} \)\( _{\mu \nu } \)(k) = G\( ^{0L} \)\( _{\mu \nu } \)(k) + i\( \int  \)\( _{_{0}} \)\( ^{1} \)d\( \kappa  \)
{[}k\( _{\mu } \)\( \widetilde{G^{0M}(k,\kappa )} \)(-ik\( ^{\sigma } \)-\( \eta ^{\sigma }) \)\( \widetilde{G^{0M}_{\sigma \nu }(k,\kappa )} \)+
(\( \mu  \),k) \( \leftrightarrow  \) (v,-k) {]}~~~ ~~ ~~~(7.1)

Here,\( \widetilde{G^{0M}(k,\kappa )} \) and \( \widetilde{G^{0M}_{\sigma \nu }(k,\kappa )} \)are
the propagators arising from the mixed gauge action together with O(\( \epsilon  \))
terms viz.S\( _{eff} \)\( ^{M}[ \)\( \phi  \),\( \kappa ] \) -i\( \epsilon  \)\( \int  \)d\( ^{4}x \)(\( \frac{1}{2} \)AA
-\( \overline{c} \) c).This can be integrated {[} 2,3{]} to yield an expression
for G\( ^{0A} \)\( _{\mu \nu } \)(k).

At first sight the expression for G\( ^{0A} \)\( _{\mu \nu } \)(k) appears
hopelessly complicated; but as discussed below,a much much simpler effective
expression can be arrived at from it.G\( ^{0A} \)\( _{\mu \nu } \)(k) reads
{[}2,3{]}:

G\( ^{0A} \)\( _{\mu \nu } \)(k)=G\( ^{0L} \)\( _{\mu \nu } \)(k)+ {[}(k\( _{\mu } \)k\( _{\nu } \)\( \Sigma _{1}+\eta _{\mu }k_{\nu } \)\( \Sigma  \)\( _{2} \))
ln \( \Sigma _{3} \) + (k\( \rightarrow  \)-k,\( \mu \rightarrow \nu  \))
~~ ~~ ~~~~ ~~ ~~~(7.2) 

where,

\( \Sigma  \)\( _{1} \)\( \equiv  \)\( \frac{-(k^{2}-i\eta .k)[\frac{\eta .k+i\eta ^{2}}{k^{2}-i\eta .k}+i\lambda -\frac{(1-\lambda )\eta .k}{k^{2}+i\epsilon }]}{\epsilon \Sigma } \);

\( \Sigma  \)\( _{2} \)\( \equiv  \)\( \frac{-(k^{2}-i\eta .k)[-(\frac{i\eta .k+k^{2}}{k^{2}-i\eta .k})+1-\frac{(1-\lambda )i\epsilon }{k^{2}+i\epsilon }]}{\epsilon \Sigma } \);

\( \Sigma  \)\( _{3} \)\( \equiv  \)\( \frac{-i(\eta .k+\epsilon )(k^{2}+i\epsilon \lambda )}{(k^{2}+i\epsilon )(-i\epsilon \lambda -\sqrt{k^{4}-(k^{2}+i\epsilon \lambda )[k^{2}+\frac{(\eta .k)^{2}+i\epsilon \eta ^{2}}{k^{2}+i\epsilon }]}} \);

with

\( \Sigma  \)\( \equiv  \)\{(1-\( \lambda  \)){[}(\( \eta  \).k)\( ^{2} \)+
2ik\( ^{2}\eta .k]+i\epsilon  \)k\( ^{2}(1-2\lambda ) \) + \( \lambda  \)(k\( ^{2}+i\epsilon )^{2}+\eta ^{2}(k^{2}+i\epsilon  \))+\( \epsilon ^{2}\lambda \} \)~
~~ ~~~~ ~~ ~~~~ ~~ ~~~(7.3)

~

These \( \Sigma  \)'s are somewhat complicated functions such that the propagator
reduces to the usual propagator away from \( \eta  \).k = 0.The above complicated
structure is not important in itself; but only its behavior as \( \epsilon  \)
--> 0.In Ref. 2,3 we have established a procedure for extracting the effective
term in this limit. For \( \eta _{0}\neq  \)0,it can be effectively replaced
by{[}2,3{]}:

{[}1{]} An integral over the contour {[}2,3{]} in the complex k\( _{0}-plane \)
that passes below\footnote{
An equivalent treatment where the contour goes \emph{above} the singularity
can also be given. 
} the singular point \( \eta  \).k=0, over a semi-circle of radius >\textcompwordmark{}>
\( \sqrt{\epsilon } \) ; so that we may replace here the propagator by its
naive expression;

{[}2{]}And a term\footnote{
There was an error in the results of Ref.2,3;please see errata. 
},\footnote{
We have set \( \eta _{0}=1 \) in the following and are assuming k\( ^{2}\neq 0;. \)and
\( \eta  \)\( ^{2} \)\( \neq  \) 0.We may take light-cone gauge limit at
the end however. 
} :

D\( _{\mu \nu }^{extra}(k)=\delta ( \)k\( _{0}-\frac{1}{2}\sqrt{-i\epsilon \eta ^{2}}-\mathbf{k}.\eta ) \)
\{k\( _{\mu } \)k\( _{\nu } \){[}i\( \sqrt{\frac{i\eta ^{2}}{\epsilon }} \)
+\( \frac{\eta ^{2}}{(\eta ^{2}+i\epsilon )} \){]} \( \bullet  \)\( \frac{2\pi \eta ^{2}}{[\mathbf{k}^{2}-(\eta .\mathbf{k})^{2}](\eta ^{2}+i\epsilon )} \)

~ ~~~~ ~~ ~~~~ ~~ ~~~~ ~~ ~~~(+{[}k\( _{\mu }\eta _{\nu }+k_{\nu }\eta _{\mu }] \)\( \frac{-i\pi \eta ^{2}}{[\mathbf{k}^{2}-(\eta .\mathbf{k})^{2}](\eta ^{2}+i\epsilon )} \)\}~
~~ ~~~~ ~~ ~~~~ ~~ ~~~~ ~~ ~~~(7.4)

We note in passing that the derivation we followed requires that we keep \( \eta ^{2} \)\( \neq  \)0
in the intermediate stages of our calculations.We can however obtain light-cone
gauge results as a limiting case in the final form. We find that in this limit
{[}\( \eta ^{2} \)\( \rightarrow  \)0, \( \epsilon  \) fixed{]}, the extra
terms D\( _{\mu \nu }^{extra}(k) \) vanish.

\section{APPLICATION TO THE WILSON LOOP AND THE THERMAL LOOP}

~

The Wilson Loop is : W{[}L{]} = < T P exp i \( \oint  \)\( _{_{C}} \)A\( _{\mu } \)dx\( ^{\mu } \)
>~ ~~ ~~~~ ~~ ~~~~ ~~ ~~~~ ~~ ~~~~ ~~ ~~~(8.1)

The preservation of the Wilson Loop {[}i.e. its having the same value for axial
and the Lorentz Gauges{]} has been taken as an important test of a prescription.For
example, the principal value prescription has been found wanting this test at
the O{[}g\( ^{4} \){]} level. For the L-M prescription, the result has been
verified to O{[}g\( ^{4} \){]} {[}15,16{]} .

Since, in our treatment, the expectation values of the gauge-invariant observables
is always preserved under the FFBRS that takes one from the Axial to the Lorentz
gauges, the formal proof that our procedure, of which the axial propagator expression
is a special element, automatically preserves the Wilson Loop.

~In reference 4, we do not rest at this formal argument:we verify it to O{[}g\( ^{4}] \).In
doing so we find it useful to make a connection with the work of Cheng \& Tsai{[}18{]}.
Using the argument in this work,we can in fact give a simple argument~ that
verifies the above result to O{[}g\( ^{4}] \) for axial gauges {[}\( \eta ^{2} \)
< 0{]} for arbitrary loops{[} not necessarily planar{]} and in other cases {[}i.e.
\( \eta ^{2} \)\( \geq  \)0{]} for a large class of loops.

The periodic Wilson Loop {[}Thermal loop{]} is another test of a prescription.It
is a rectangular Wilson loop in Euclidean space in the {[}x,\( \tau  \){]}
plane with side along \( \tau  \) axis from 0 to 1/\( \beta . \)The numerical
result for the loop to ~O{[}g\( ^{2} \){]} is easy to evaluate explicitly {[}20{]}.
This result for the loop to O{[}g\( ^{2} \){]} reads\footnote{
There were unfortunate trivial typographical errors in {[}4{]} related to D\( _{00} \)
{[}k\( _{0} \)=0, \textbf{k{]}} in the result for the thermal loop,see erratum{[}4{]}. 
} :

~ W\( _{R} \) = 1+\( \frac{(N^{2}_{c}-1)g^{2}}{2N_{c}T} \)\( \int  \)\( \frac{d^{3}k}{(2\pi )^{3}} \){[}1-\( \cos  \)(k.R){]}D\( _{00} \)
{[}k\( _{0} \)=0, k{]}~ ~~ ~~~~ ~~ ~~~~ ~~ ~~~~ ~~ ~~~~ ~~ ~~~(8.2)

The above quantity, depends on D\( _{00} \) {[}k\( _{0} \)=0, k{]}. ~ ~It
is easy to verify from the expression (8.2), that this quantity has the same
value as in Lorentz gauges.Thus, the thermal Wilson loop is preserved to O{[}g\( ^{2} \){]}.Thus,
Our prescription satisfies this test also.

\section{SUMMARY AND COMMENTS:}

{[}A{]} We generated a field-dependent BRS {[}FFBRS{]} transformation that connects
different classes of gauges {[}the procedure is general enough{]};this transformation
preserves the vacuum expectation value of gauge-invariant observables;

{[}B{]} We applied it to the connection between the Lorentz and the axial-type
gauges ;

{[}C{]}We gave a compact result that can relate the Green's functions in the
two classes of gauges;

{[}D{]}We applied the procedure to the question of the treatment of the axial
gauge pole prescription;

{[}E{]}We showed that the Wilson Loop is preserved under our transformation
and verified the result to O{[}g\( ^{4} \){]}.

Recently,we have generalized these results to an arbitrary pair of gauges for
the FPEA and to the planar gauges{[}17{]}.

\section{FUTURE PLANS}

Finally we summarize the future plans of development along these lines of work.

{[}1{]}Use the understanding developed to address to the existing problems with
the axial and Coulomb gauges .

{[}2{]}Use the relation (5.9) to study Green's functions (both primary and of
simple gauge-invariant operators ) in two sets of gauges; and their renormalization
properties and anomalous dimensions.

{[}3{]}Develop formal techniques for establishing gauge-independence of arbitrary
observables such as observable anomalous dimensions and cross-sections.

Work along these lines is in progress.

~Acknowledgement:

The work in part was supported by grant for the DST project No. DST/PHT/19990170.

\end{document}